\documentclass[cernpreprint,english,a4paper]{na61doc}
\usepackage{blindtext}

\usepackage{colortbl}
\definecolor{darkred}{rgb}{0.5,0,0}
\definecolor{darkblue}{rgb}{0,0,0.5}
\definecolor{firebrick}{rgb}{0.75,0.125,0.125}
\definecolor{darkgreen}{rgb}{0,0.5,0}

\usepackage[colorlinks=true,linkcolor=firebrick,citecolor=darkgreen,urlcolor=darkblue]{hyperref}

\usepackage{graphicx}
\usepackage{amsmath,amssymb,amsfonts}
\usepackage{bbding}
\usepackage{wasysym}%'Multistrangeness_DP(2).tex'
\usepackage{afterpage}
\usepackage{mathpazo}
\usepackage{overpic}
\usepackage{multicol}
\usepackage{subfigure}
\usepackage{xspace}
\usepackage{rviewport}
\usepackage{colortbl}
\usepackage[toc,page]{appendix}
\usepackage{enumerate}

\usepackage{blindtext}

\usepackage{tikz}

\usepackage{lineno}
\usepackage{colortbl}
\usepackage{enumerate}
\usepackage{listings}
\usepackage{appendix}
\usepackage[final]{pdfpages}
\usepackage[Symbol]{upgreek}
\usepackage{makecell}
\definecolor{kOrange+8}{RGB}{255,102,51}
\definecolor{kBlue}{RGB}{0,0,204}
\definecolor{kRed+2}{RGB}{153,0,0}
\definecolor{kGreen}{RGB}{0,153,0}
\definecolor{kAuAu}{RGB}{255,153,153}
\definecolor{kRed}{RGB}{204,0,0}

\graphicspath{{plots/}}
\definecolor{darkred}{rgb}{0.5,0,0}
\definecolor{darkblue}{rgb}{0,0,0.5}
\definecolor{firebrick}{rgb}{0.75,0.125,0.125}
\definecolor{darkgreen}{rgb}{0,0.5,0}

\definecolor{colorArSc}{RGB}{255,102,51}
\definecolor{colorNN}{RGB}{0,0,204}
\definecolor{colorNNWorld}{RGB}{153,153,255}
\definecolor{colorBeBe}{RGB}{0,153,0}
\definecolor{colorXeLa}{RGB}{204,0,255}
\definecolor{colorPbPb}{RGB}{204,0,0}
\definecolor{colorAuAu}{RGB}{204,0,0}

\usepackage[colorlinks=true,linkcolor=firebrick,citecolor=darkgreen,urlcolor=darkblue]{hyperref}
\ShineTitle{Proposal from the \NASixtyOne Collaboration \\ for update of European Strategy for Particle Physics}
\ShineAbstract{
{\footnotesize
Building on the current program's success and driven by new physics challenges, the \NASixtyOne Collaboration proposes to continue measuring hadron production properties in reactions induced by hadron and ion beams after CERN Long Shutdown~3. These measurements are of significant interest to the heavy-ion, cosmic-ray, and neutrino physics communities and will focus on:
\begin{enumerate}[(i)]
\item Investigating hadron production in the light-ion systems to explore the diagram of high-energy nuclear collisions, and to obtain new insight into the unexpected violation of isospin (flavor) symmetry recently observed by the experiment;
\item Measuring charm-anticharm correlations to gain unique insights into the production locality of charm and anticharm quark pairs;
\item Examining strangeness and multi-strangeness production to improve our understanding of the early Universe's evolution and neutron star formation;
\item Measuring cross sections relevant for cosmic-ray measurements, significantly boosting searches for new physics in our Galaxy;
\item Conducting hadron production measurements with proton, pion, and kaon beams for neutrino physics, enhancing the precision of hadron production data needed for initial neutrino flux predictions in neutrino oscillation experiments;
\item Measuring hadron production processes relevant for understanding the flux of atmospheric neutrinos, as well as neutrinos and muons from spallation sources. 
\end{enumerate}
To achieve these objectives, a detector upgrade and a beam upgrade are required, with data-taking planned for the period 2029–2032 and beyond. 
}}

\begin{document}
%%%%%%% units
\newcommand{\kg}{\ensuremath{\mbox{kg}}\xspace}
\newcommand{\eV}{\ensuremath{\mbox{e\kern-0.1em V}}\xspace}
\newcommand{\GeV}{\ensuremath{\mbox{Ge\kern-0.1em V}}\xspace}
\newcommand{\MeV}{\ensuremath{\mbox{Me\kern-0.1em V}}\xspace}
\newcommand{\byc}{\kern-0.1em/\kern-0.1em c}
\newcommand{\GeVc}{\ensuremath{\mbox{Ge\kern-0.1em V}\byc}\xspace}
\newcommand{\GeVcc}{\ensuremath{\mbox{Ge\kern-0.1em V}\byc^2}\xspace}
\newcommand{\MeVcc}{\ensuremath{\mbox{Me\kern-0.1em V}\byc^2}\xspace}
\newcommand{\AGeV}{\ensuremath{A\,\mbox{Ge\kern-0.1em V}}\xspace}
\newcommand{\AGeVc}{\ensuremath{A\,\mbox{Ge\kern-0.1em V}\byc}\xspace}
\newcommand{\MeVc}{\ensuremath{\mbox{Me\kern-0.1em V}\byc}\xspace}
\newcommand{\T}{\ensuremath{\mbox{T}}\xspace}
\newcommand{\cmsq}{\ensuremath{\mbox{cm}^2}\xspace}
\newcommand{\msq}{\ensuremath{\mbox{m}^2}\xspace}
\newcommand{\cm}{\ensuremath{\mbox{cm}}\xspace}
\newcommand{\mm}{\ensuremath{\mbox{mm}}\xspace}
\newcommand{\micron}{\ensuremath{\mu\mbox{m}}\xspace}
\newcommand{\mrad}{\ensuremath{\mbox{mrad}}\xspace}
\newcommand{\ns}{\ensuremath{\mbox{ns}}\xspace}
\newcommand{\m}{\ensuremath{\mbox{m}}\xspace}
\newcommand{\s}{\ensuremath{\mbox{s}}\xspace}
\newcommand{\ms}{\ensuremath{\mbox{ms}}\xspace}
\newcommand{\ps}{\ensuremath{\mbox{ps}}\xspace}
\newcommand{\dd}{\ensuremath{{\textrm d}}\xspace}
\newcommand{\dedx}{\ensuremath{\dd E\!~/~\!\dd x}\xspace}
\newcommand{\tofdedx}{\ensuremath{\textup{\emph{tof}}-\dd E/\dd x}\xspace}
\newcommand{\tof}{\ensuremath{\textup{\emph{tof}}}\xspace}
\newcommand{\pt}{\ensuremath{p_{\textrm T}}\xspace}
\newcommand{\PT}{\ensuremath{P_\textup{T}}\xspace}
\newcommand{\mt}{\ensuremath{m_{\textrm T}}\xspace}
\newcommand{\y}{\ensuremath{{y}}\xspace}

%particles
\newcommand{\pbar}{\ensuremath{\overline{\textit{p}}}\xspace}
\newcommand{\nbar}{\ensuremath{\overline{\textit{n}}}}
\newcommand{\dbar}{\ensuremath{\overline{\textup{d}}}}
\newcommand{\pim}{\ensuremath{\pi^-}\xspace}
\newcommand{\pip}{\ensuremath{\pi^+}\xspace}
\newcommand{\km}{\ensuremath{\textit{K}^-}\xspace}
\newcommand{\kp}{\ensuremath{\textit{K}^+}\xspace}
\newcommand{\hm}{\ensuremath{\textit{h}^-}\xspace}
\newcommand{\Xim}{\ensuremath{\Xi^-}\xspace}
\newcommand{\Xip}{\ensuremath{\overline{\Xi}^+}\xspace}

%reactions
\newcommand{\pp}{\mbox{\textit{p+p}}\xspace}
\newcommand{\pA}{\mbox{\textit{p}+A}\xspace}
\newcommand{\NN}{\mbox{\textit{N+N}}\xspace}
\newcommand{\snn}{\ensuremath{\sqrt{s_{\mathrm{NN}}}}\xspace}

% inverse hyperbolic functions
%\DeclareMathOperator{\acosh}{acosh}
%\DeclareMathOperator{\asinh}{asinh}
%\DeclareMathOperator{\atanh}{atanh}

%%%%%%%%%%%%% some software programs and generators
%----- NA61 software
\def\Offline{\mbox{$\overline{\text%
{Off}}$\hspace{.05em}\raisebox{.4ex}{\underline{line}}}\xspace}
\def\SHOE{\mbox{SHO\hspace{-1.34ex}\raisebox{0.2ex}{\color{green}\textasteriskcentered}\hspace{0.25ex}E}\xspace}
\def\DSHACK{\mbox{DS\hspace{0.15ex}$\hbar$ACK}\xspace}
\DeclareRobustCommand{\SHINE}{\mbox{\textsc{S\hspace{.05em}\raisebox{.4ex}{\underline{hine}}}}\xspace} %DeclareRobustCommand allows this to work in caption
%\def\SHINE{\textsc{Shine}\xspace}
%----- event generators
\def\Glissando{\textsc{Glissando}\xspace}
\newcommand{\FlukaLong}{{\scshape Fluka2008}\xspace}
\newcommand{\FlukaEleven}{{\scshape Fluka2011}\xspace}
\newcommand{\Fluka}{{\scshape Fluka}\xspace}
\newcommand{\UrqmdLong}{{\scshape U}r{\scshape qmd1.3.1}\xspace}
\newcommand{\Urqmd}{{\scshape U}r{\scshape qmd}\xspace}
\newcommand{\GheishaLong}{{\scshape Gheisha2002}\xspace}
\newcommand{\GheishaOld}{{\scshape Gheisha600}\xspace}
\newcommand{\Gheisha}{{\scshape Gheisha}\xspace}
\newcommand{\Corsika}{{\scshape Corsika}\xspace}
\newcommand{\Venus}{{\scshape Venus}\xspace}
\newcommand{\VenusLong}{{\scshape Venus4.12}\xspace}
\newcommand{\GiBUU}{{\scshape GiBUU}\xspace}
\newcommand{\GiBUULong}{{\scshape GiBUU1.6}\xspace}
\newcommand{\FlukaNewLong}{{\scshape Fluka2011.2\_17}\xspace}
\newcommand{\Root}{{\scshape Root}\xspace}
\newcommand{\Geant}{{\scshape Geant}\xspace}
\newcommand{\GeantThree}{{\scshape Geant3}\xspace}
\newcommand{\GeantFour}{{\scshape Geant4}\xspace}
\newcommand{\QGSJet}{{\scshape QGSJet}\xspace}
\newcommand{\DPMJet}{{\scshape DPMJet}\xspace}
\newcommand{\Epos}{{\scshape Epos}\xspace}
\newcommand{\EposLong}{{\scshape Epos1.99}\xspace}
\newcommand{\QGSJetLong}{{\scshape QGSJetII-04}\xspace}
\newcommand{\DPMJetLong}{{\scshape DPMJet3.06}\xspace}
\newcommand{\SibyllLong}{{\scshape Sibyll2.1}\xspace}
\newcommand{\EposLHCLong}{{\scshape EposLHC}\xspace}
\newcommand{\Hsd}{{\scshape Hsd}\xspace}
\newcommand{\Ampt}{{\scshape Ampt}\xspace}
\newcommand{\Hijing}{{\scshape Hijing}\xspace}
\newcommand{\PHSD}{{\scshape Phsd}\xspace}
\newcommand{\SmashModel}{{\scshape Smash}\xspace}
\newcommand{\FTFPBERT}{{\scshape Ftfp-Bert}\xspace}
\newcommand{\Pythia}{{\scshape Pythia}\xspace}
\newcommand{\SMES}{{\scshape Smes}\xspace}
\newcommand{\ALCOR}{{\scshape Alcor}\xspace}

%%%%%%%%%%%%%%%%%%%%%%%% misc
\def\red#1{{\color{red}#1}}
\def\blue#1{{\color{blue}#1}}
\def\avg#1{\langle{#1}\rangle}
\def\sci#1#2{#1\!\times\!10^{#2}}
\newcommand{\Fi}[1]{Fig.~\ref{#1}}
\newcommand{\CernVM}{\textsc{Cern\-\kern-0.05emVM}\xspace}
\newcommand{\coordinate}[1]{{\fontfamily{lmss}\selectfont#1}}
%%%%%%%%%%%%%%%%%%%%%%%% for cosmic section
\def \pions{$\pi^\pm$\xspace}
\def \kaons{K$^\pm$\xspace}
\def \proton{p\xspace}
\def \antiproton{$\bar{\text{p}}$\xspace}
\def \protons{p($\bar{\text{p}}$)\xspace}
\def \lamb{$\Lambda$\xspace}
\def \antilamb{$\bar{\Lambda}$\xspace}
\def \lambs{$\Lambda(\bar{\Lambda})$\xspace}
\def \kzeros{K$_{S}^{0}$\xspace}
\def \ncl{$N_{\mathrm{cl}}$\xspace}
\def \shine{NA61/SHINE\xspace}
\def \pipi{$\pi^+\pi^-$\xspace}
\def \vzero{$V^0$\xspace}
\def \kaonstar{K$^{*0}$\xspace}
\def \rhozero{$\rho^{0}$\xspace}
\newcommand{\pT}{\ensuremath{p_\text{T}}\xspace}
\newcommand{\TeV}{\ensuremath{\mbox{Te\kern-0.1em V}}\xspace}

\maketitle

\newpage
%\tableofcontents

\section{Scientific context}
%{\itshape \color{blue} Coordinator: Andrzej}

\NASixtyOne~\cite{NA61:2014lfx} is a multipurpose, fixed-target experiment to study proton-proton, hadron-nucleus, and nucleus-nucleus collisions at the CERN Super Proton Synchrotron (SPS). The CERN Research Board approved the experiment in 2007, based on the request of heavy-ion, neutrino, and cosmic-ray communities. Its research program includes: 
\begin{itemize}
\item 
Measurements for physics of strong interactions; 
\item 
Measurements of hadron cross sections to reduce uncertainties in accelerator neutrino flux;
\item
Studies of hadron production and nuclear fragmentation to improve understanding of cosmic-ray propagation in the Galaxy. 
\end{itemize}

As the only detector of this type operating at the SPS, \NASixtyOne has the unique capability for measurements over a versatile set of beams and targets in the collision center-of-mass system (c.m.s.) energy regime $5<\snn<17$~GeV. The presently accumulated data set includes $p$+$p$, $p$+C, $\pi$+C, $K$+C, $p$+Pb, Be+Be, C+C, Ar+Sc, Xe+La, and Pb+Pb reactions, recorded at up to six beam momenta per reaction. It also includes large data samples obtained with replicas of targets of neutrino experiments T2K, NuMI, and LBNF/DUNE. A significant advantage of \NASixtyOne over collider experiments is its extended coverage of phase space available for hadron production. This includes the entire forward c.m.s.~hemisphere for charged hadrons and a large part of the entire reaction sphere for neutral hadrons. 

In its input to the previous update of the European Strategy for Particle Physics (ESPP), the Collaboration proposed a detailed research program, which is now nearing completion. The most recent achievements of this program include:
\begin{enumerate}[(i)]
\item
  The first-ever direct measurement of open charm production in nucleus-nucleus collisions at SPS energies~\cite{SQM2024_Anastasia};
\item
 The observation of an unexpected excess of charged over neutral $K$ meson production in Ar+Sc collisions at $\snn=11.9$~GeV, interpreted as evidence for a violation of isospin (flavor) symmetry beyond known effects~\cite{NA61SHINE:2023azp};
\item
  The achievement of a 2D scan of hadron production in $p$+$p$ and nucleus-nucleus collisions at six beam momenta per reaction, bringing insight into the complicated changeover from confined to deconfined matter as a function of collision energy and system size~\cite{Rybicki:2024mfy};
\item Data collections with a T2K replica target (2022) and LBNF prototype target (2024, 2025) that achieved over a hundred million triggered events each, which is more than ten times higher statistics compared to data sets collected before the upgrade (LS2)~\cite{ABGRALL201399}; 
\item Carbon fragmentation measurements at 13.5\AGeVc~\cite{NA61SHINE:2024rzv} were performed, which are crucial for precisely determining cosmic-ray transport parameters in the Galaxy; 
\item First deuteron production measurements in $p$+$p$ in the most relevant energy range for building an astrophysical background model of cosmic antideuteron production.
  \end{enumerate}
The upgraded \NASixtyOne detector records high-statistics Pb+Pb data, bringing the first-ever differential measurement of open charm production in heavy-ion collisions close to the threshold to be completed in 2026.
%At the same time, the detector provides another substantial set of reference data to improve the precision of neutrino experiments and new measurements for cosmic-ray physics programs.

Our present input summarizes $\bullet$ the immediate physics goals of the \NASixtyOne Collaboration in terms of strong-interaction, neutrino, and cosmic-ray physics (Sec. 2); $\bullet$ the planned upgrades of the detector and beamline (Sec. 3); $\bullet$ readiness and expected challenges (Sec. 4); and $\bullet$~our recommendations for support by the ESPP (Sec. 5).

\section{Objectives}
The \NASixtyOne experiment plans to continue measurements after CERN Long Shutdown 3 during Run 4. The primary objectives are to explore transitions to quark-gluon plasma (QGP) in light-ion systems through comprehensive, large-acceptance hadron measurements and to study hadron production relevant to neutrino and cosmic-ray physics in the SPS energy range.

The main goals of measurements for the physics of strong interactions are:
\begin{itemize}
    \item Investigating hadron production in the light-ion systems to explore the diagram of high-energy nuclear collisions, and to obtain new insight into the unexpected violation of isospin (flavor) symmetry recently observed by the experiment;
    \item Measuring charm-anticharm correlations to provide unique insights into the production locality of charm and anticharm quark pairs;
    \item Examining strangeness and multi-strangeness production to enhance our understanding of the early Universe's evolution and neutron star formation.
\end{itemize}
%Investigating hadron production in medium-sized collision systems to explore the phase diagram of strongly interacting matter.
%Measuring charm–anti-charm correlations to provide unique insights into the production locality of charm and anti-charm quark pairs.
%Examining strangeness and multistrangeness production to enhance our understanding of the early Universe's evolution and neutron star formation.

The objectives related to neutrino and cosmic-ray physics are as follows:
\begin{itemize}
    \item Studies of hadron production on nuclear targets using a new very-low-energy beam;
    \item Measurements with new replica targets for the LBNF/DUNE and Hyper-Kamiokande neutrino experiments;
    \item Determination of antiproton cross sections and parameters of coalescence models for antinuclei;
    \item High-precision measurements of inelastic and production cross sections necessary for Galactic cosmic-ray data interpretation. 
\end{itemize}
\subsection{Hadron production in light-ion systems}
%{\itshape \color{blue} Coordinator: Maja}

Assuming a statistical description of strongly interacting matter, the phase diagram of the latter can be described by the dominance of different phases. One can distinguish the hadron phase at low temperature $T$ and the QGP phase at high $T$. At higher values of baryo-chemical potential ($\mu_B$), other exotic forms of matter may be present. At sufficiently high temperatures and energy densities, the system undergoes a phase transition from the hadron to the QGP phase. One of the signatures of such transition, predicted by the Statistical Model of the Early Stage (SMES)~\cite{Poberezhnyuk:2015wea}, is the ratio of positively charged kaons to positively charged pions ($K^{+}/\pi^{+}$) which is related to the strangeness-to-entropy ratio. This ratio should increase with energy in the hadron phase, reach a maximum at the phase transition, and then decrease and approach saturation.
\begin{figure}[ht]
    \centering
    \includegraphics[width=0.35\linewidth]{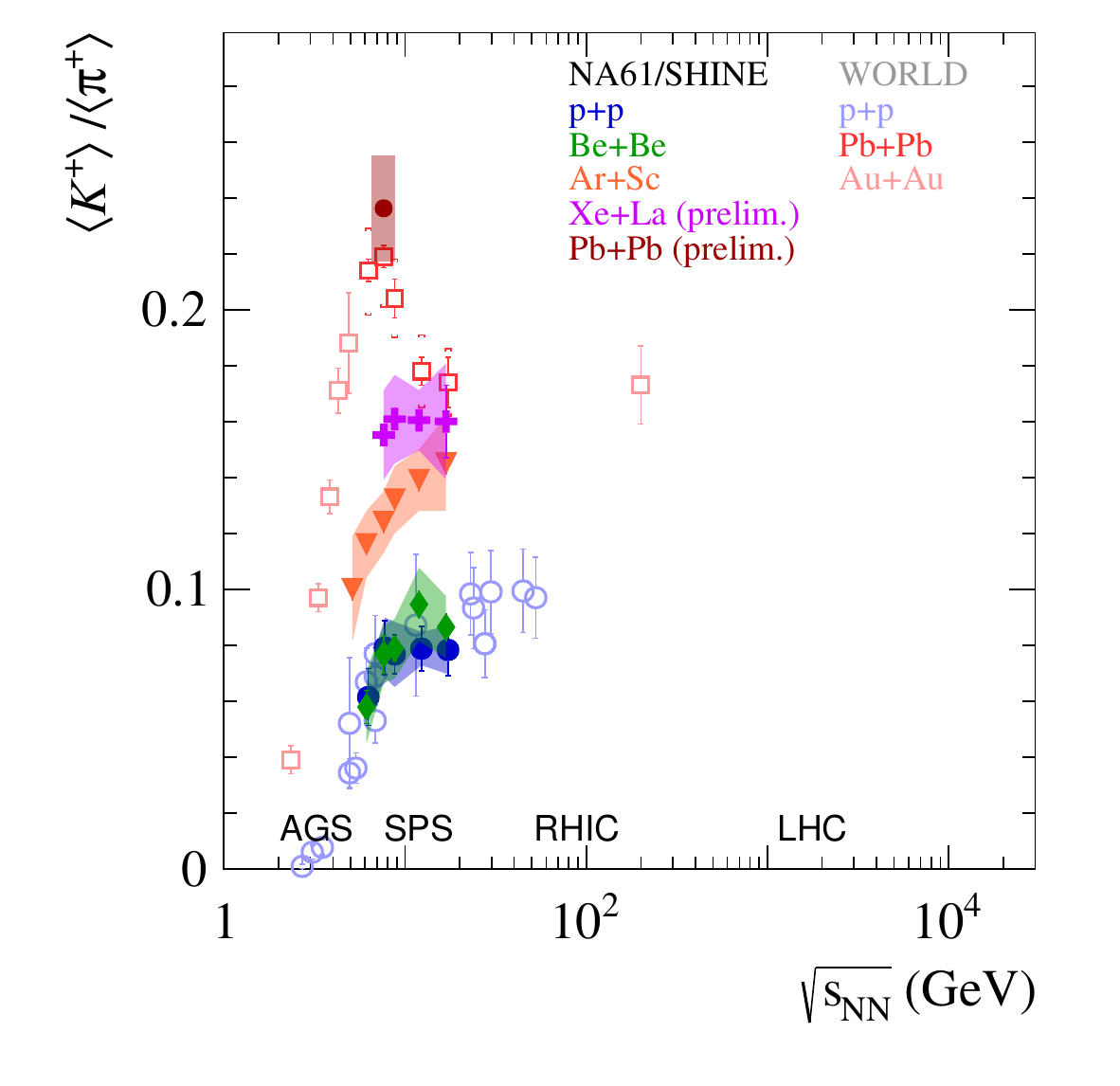}
    \includegraphics[width=0.35\linewidth]{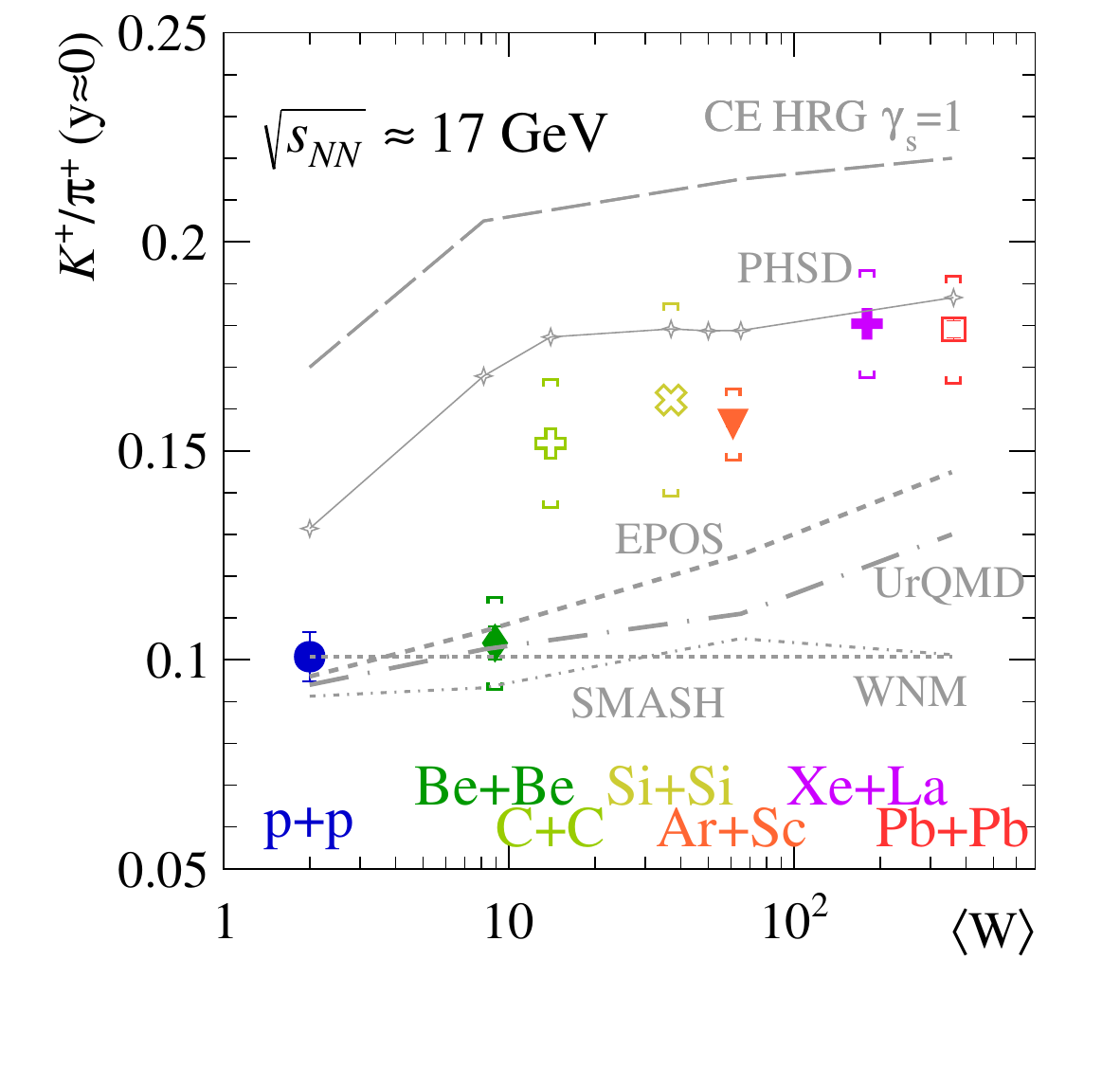}
    \includegraphics[width=0.55\linewidth]{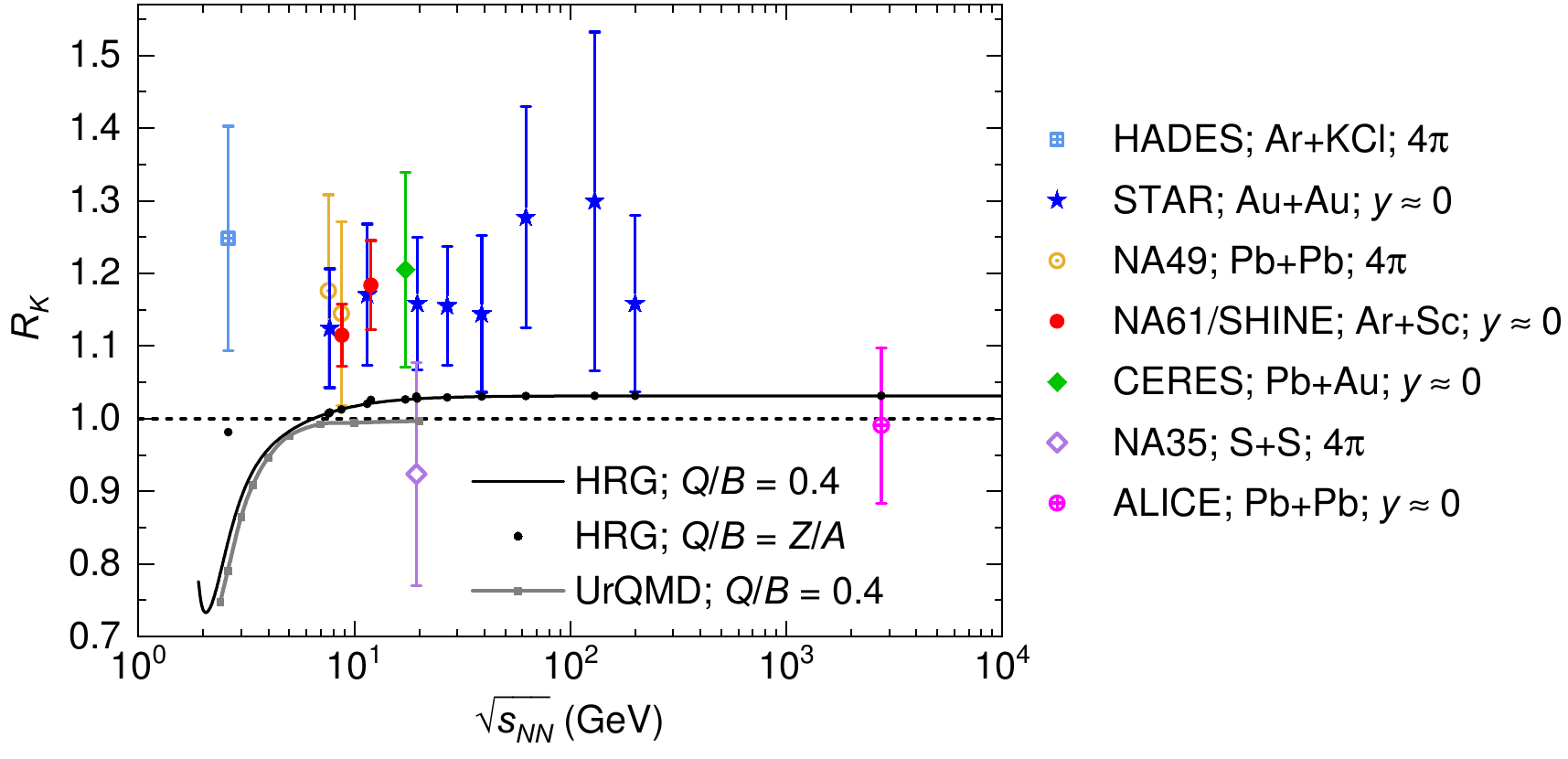}
    \caption{\textit{Top left}: Energy dependence of $\langle K^{+}\rangle/\langle \pi^{+}\rangle$ mean multiplicity ratio. \textit{Top right}: System size dependence of mid-rapidity $ K^{+}/\pi^{+}$ ratio at $\snn=16.8~(17.3)$ GeV. \textit{Bottom}: Energy dependence of $R_{K}$ (see Eq.~\ref{Eq:rk}).}
    \label{fig:horn}
\end{figure}
This was indeed first observed (see the top left panel of Fig.~\ref{fig:horn}) by the NA49 experiment in Pb+Pb collisions in comparison to \pp interactions, where no such structure could be seen. The system size and energy scan of \NASixtyOne aimed to further study the hadron production mechanism at intermediate system sizes. The results obtained within the scan in \pp, Be+Be, Ar+Sc, and Xe+La indicate a transition of hadron production mechanism between light systems (\pp and Be+Be) and heavy ones (Ar+Sc, Xe+La, and Pb+Pb).

The most popular models describing the hadron production process are the creation and decay of resonances or strings and the formation and hadronization of QGP.  
The measurement indicates that the applicability domains of present models in the space of laboratory-controlled parameters (collision energy and nuclear mass number of colliding nuclei) have not been well-established~\cite{Andronov:2022cna}. 

For equal quark masses and in the absence of other interactions, QCD assumes that interactions are independent of quark type (flavor). This feature is called flavor symmetry. It reduces to isospin symmetry when only up and down quarks are considered. During an ion+ion collision, many new particles are the lightest mesons -- pions and kaons, built up from up (\textit{u}), down (\textit{d}), and strange (\textit{s}) quarks. The effective masses of up and down quark are $m_{u}=2.16(7)$ MeV and $m_{d}=4.70(7)$ MeV. 
These are not equal but much smaller than the QCD scale, $\Lambda_{\mathrm{QCD}}$. Hence the isospin symmetry breaking is small, as confirmed by the mass ratio of pions or kaons, $(m_{\pi^{+}}-m_{\pi^{0}})/(m_{\pi^{+}}+m_{\pi^{0}})\backsimeq 0.017$; for $K^{+}$ and $K^{0}$ it is $-$0.004. 
The elastic cross sections for pion-pion, pion-nucleon, and nucleon-nucleon scattering support the isospin symmetry. One of the specific isospin transformations is an inversion of the third component of the isospin. It is equivalent to swapping $u\leftrightarrow d$ and in hadronic case implies swapping $p\leftrightarrow n$, $\pi^{+}\leftrightarrow \pi^{-}$, $K^{+}\leftrightarrow K^{0}$, $\overline{K}^0\leftrightarrow K^{-}$, etc. 
In a nucleus-nucleus collision where the nuclei have an equal number of protons and neutrons, strong interactions are expected to be invariant under charge symmetry transformation for every nucleus and hadron in the initial and final state. The invariance implies that the mean number of charge transformation-related hadrons, such as $K^{+}$ and $K^{0}$ as well as $\overline{K}^{0}$ and $K^{-}$, coincide. It should be underlined that models predict only properties of ensembles, so one needs to consider averaged quantities. Thus, the ratio of charged to neutral kaons in interactions of ions with $n=p$ should be rather:
\begin{equation}
    R_{K}=\frac{\langle K^{+}\rangle + \langle K^{-}\rangle}{2\langle K^{0}_{S}\rangle}=1\, .
    \label{Eq:rk}
\end{equation}
\NASixtyOne measured the unexpectedly large value of $R_{K} = 1.184(61)$ 
in the 10$\%$ most central Ar+Sc collisions at $\snn=11.9$~GeV~\cite{NA61SHINE:2023azp}. The measurement, as well as comparison to other experimental results and the new preliminary \NASixtyOne point at \snn = 8.8~GeV~\cite{QM2025_YB}, is shown in the bottom panel of Fig.~\ref{fig:horn}. The ratio $R_{K}$ is systematically larger than one across the available energy range, but at the LHC energy.  
The results were compared with models' predictions, calculated for reactions corresponding to the experimental data, which generally have $Q/B < 1/2$, where $Q$ and $B$ represent the total net charge and net baryon number in the system. The models take into account this and other known isospin symmetry-breaking effects. They fail to describe the experimental data.

To uncover the nature of the transition from light to heavy systems and to study the unexpected large violation of isospin, \NASixtyOne proposed a new set of measurements with light ions ($A<40$) at three beam momenta:
\begin{itemize}

\item the top SPS momentum, 150\AGeVc, to locate the rapid changeover between strings and QGP production,

\item  the intermediate SPS momentum, 30\AGeVc, to establish the change with the nuclear mass number close to the region in which all three processes may contribute: resonances, strings, and QGP,

\item the lowest SPS ion-beam momentum, 13\AGeVc, to establish a continuous change with the nuclear mass number (absence of a changeover) in the resonance domain.
\end{itemize}
The proposed systems are $^{16}$O+$^{16}$O, $^{24}$Mg+$^{24}$Mg and $^{10}$B+$^{10}$B. They all utilize optimal \NASixtyOne detector configuration~\cite{NA61/SHINEaddedndum}.

\subsection{Charm-anticharm correlations}
%{\itshape \color{blue} Coordinator: Marek}

Heavy, rarely produced charm and bottom quarks play a special role~\cite{vanHees:2004gq, Blaizot:2017ypk} in studies of quark-gluon plasma, whose properties are predominantly determined by light quarks and gluons. In particular, heavy quark correlations have a high chance of providing insight into the physics of strongly interacting matter and a considerable discovery potential.

Previously, the momentum correlations of charm and bottom hadrons have been considered for testing heavy quarks' thermalization in the hot, dense medium produced by the collisions. 
In this respect, two effects have been considered: the decrease of the initial back-to-back correlations~\cite{Zhu:2006er, Cao:2015cba} and the increase of correlations due to heavy-quark interactions with collectively flowing medium~\cite{He:2019vgs}.

Recently, it was proposed~\cite{Gazdzicki:2023niq,Gazdzicki:2025jng} that, in the case of a single charm and anticharm hadron pair production, the collective flow of matter created in heavy-ion collisions allows for testing heavy-quark production locality and transport properties.
The first results on $D$ meson production in central Xe+La collisions at 
150\AGeVc~\cite{SQM2024_Anastasia} indicates that the mean multiplicity of charm-anticharm quark pairs is below one. Using the collision number scaling, one expects about one
$c-\bar{c}$ pair in central Pb+Pb collisions at this beam momentum.
The expectation will be soon verified by the ongoing \NASixtyOne measurements~\cite{Kowalski:2916893}. This makes the measurements of charm-anticharm correlations at the CERN SPS very promising.

\begin{figure}[ht]
\begin{center}
\includegraphics[width=0.6\textwidth]{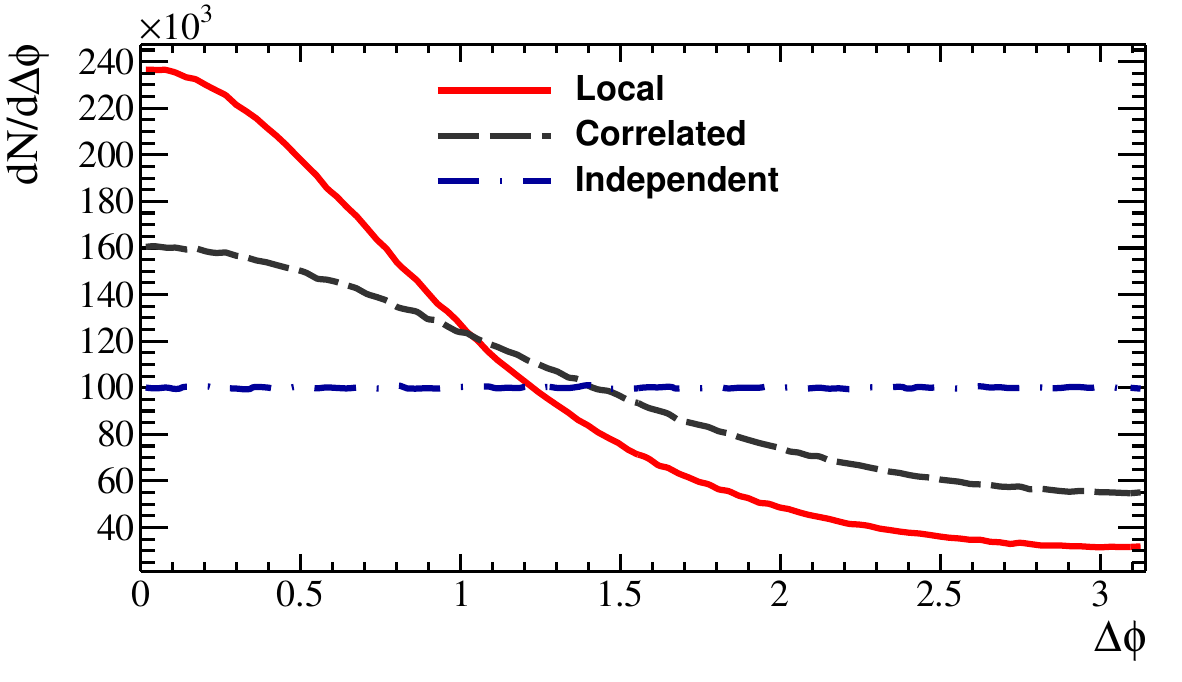} 
\end{center}
\caption{
Distribution of charm-anticharm hadron pairs 
in the difference of azimuthal angles $\Delta\phi$ simulated for central Pb+Pb collisions at 150\AGeVc for local, independent and correlated ($\sigma = 2$~fm) emission models; see Ref.~\cite{Gazdzicki:2023niq,Gazdzicki:2025jng} for detail.
}
\label{fig:Sim:Corr}
\end{figure}

The azimuthal correlations of charm and anticharm hadrons are expected to be particularly sensitive to their spatial correlations at freeze-out~\cite{Gazdzicki:2023niq,Gazdzicki:2025jng}.
Figure~\ref{fig:Sim:Corr} shows the distribution of $c-\bar{c}$ hadron pairs 
in the difference of azimuthal angles $\Delta\phi$ for local, independent, and correlated emission from the freeze-out hypersurface (see Ref.~\cite{Gazdzicki:2023niq,Gazdzicki:2025jng} for detail).

Thus, measurements of the correlation between charm and anticharm hadrons in central Pb+Pb collisions at the top CERN SPS energy should provide a unique input verifying assumptions concerning the production locality
of a charm and anticharm quark pair and their subsequent subliminal transport in the dense medium. Upgrading the \NASixtyOne detector utilizing the existing experimental technology and beam intensities at the H2 beamline of the CERN SPS should allow for the corresponding measurements as early as in Run~4 (see below).

\subsection{Strangeness and multi-strangeness production}
%{\itshape \color{blue} Coordinator: Joanna}

Strangeness production is an important tool for understanding the development
of the early Universe and the formation of neutron stars.
An important consideration, first noted by Ref.~\cite{Witten:1984rs}, is that in the quark-gluon plasma, strange quarks' admixture lowers the Fermi momentum by redistributing the available quarks across three flavors (up, down, and strange). This has significant implications for the study of strangeness in heavy-ion collisions, as it suggests that measuring strangeness production may not only probe the QGP but also provide insight into the underlying quark dynamics at high densities.

 An enhancement of strange quark yields serves as a key signal of QGP formation in heavy-ion collisions. A major discovery in this context was the horn structure in the $K^+/\pi^+$ ratio, first observed by the NA49 experiment \cite{NA49}, the predecessor of \NASixtyOne, which inherited most of its detector setup. \NASixtyOne, operating at SPS energies, is particularly well-suited for investigating the onset of deconfinement, as this energy range covers the transition region where the QGP is expected to emerge.
\NASixtyOne is performing a systematic scan of strange meson and baryon production $K^0_S$, $K^+$, $K^-$, $K^{*}(892)^0$ and $\Lambda$ in the nucleon-nucleon center-of-mass energies ranging between 5 and 17~\GeV and for different sizes of colliding nuclei -- an improvement in the detector performance allowed for a significant increase in the statistics of collected events.
We can now study the multi-strangeness production, which refers to the appearance of events with multiple lambdas, as observed already in Ar+Sc interactions.
Conditions for studying light hyperfragments and multi-hyperfragments appear to be more favorable in the SPS energy range. We plan to search for predicted but yet unobserved H-hyperfragment,  \textit{uuddss} state~\cite{ALICE:2015udw}.
The chance for the $\Lambda$ absorption in the nuclear matter is expected to
decrease with the incident energy of nuclei as predicted in the coalescence and thermal model~\cite{Reichert:2022mek}. Therefore, the signal-to-background ratio may be more favorable at SPS than at the LHC.

Beyond the established role of strangeness production, event-by-event fluctuations of charged and neutral kaon yields offer an additional perspective on potential signatures of disoriented chiral condensates (DCCs) and chiral symmetry restoration. A recent example is the study performed by the \mbox{ALICE} Collaboration on correlations between charged and neutral kaons in Pb+Pb collisions at $\snn = 2.76$~TeV~\cite{ALICE:2021fpb}, which explored possible DCC-related effects in the kaon sector. While no direct evidence for DCC formation was found, this analysis underscores the continued interest in such searches.

Beyond hadronic probes, dielectron production has been proposed as another potential signature of DCC formation. Theoretical 
studies~\cite{Kluger:1997cm, Lee:2011nj} suggest that dileptons provide an electromagnetic window into the early stages of chiral symmetry breaking. A systematic approach combining strangeness fluctuations and dilepton measurements could offer a more comprehensive understanding of the behavior of the QCD matter under extreme conditions and further test the possible manifestation of chiral symmetry restoration.

\subsection{Cosmic rays}

Cosmic-ray physics in the \GeV-to-\TeV  energy range has entered a precision era thanks to recent data from space-based experiments. However, the limited knowledge of nuclear reactions, particularly in the production of antimatter and secondary nuclei, restricts the information that can be extracted from these data, such as source properties, transport in the Galaxy, and indirect searches for particle dark matter. 
A recent review~\cite{maurin2025precisioncrosssectionsadvancingcosmicray} provides a detailed roadmap to close the most urgent gaps in cross-section data, thereby efficiently progressing on many open physics cases.
\NASixtyOne has the capabilities of making significant contributions to the interpretation of available cosmic-ray data. 

\subsubsection{Antiproton cross sections and coalescence for antinuclei}

In recent years, the AMS-02 Collaboration has reported detecting several cosmic antideuteron candidate events in the energy region of a few GeV per nucleon~\cite{2022cosp...44.2082C}.
%~\cite{2022cosp...44.2082C,antihe, antihe2, antihe3,antihe4, 2022cosp...44.2083C, tingcolloq2023}. 
It is an open question whether the explanation of these events requires an explanation invoking new physics, like dark matter annihilation or decay,
or can be attributed to known astrophysical processes.

The astrophysical background flux of antideuterons is predominantly produced through proton interactions with interstellar hydrogen and helium gas. The antideuteron production cross section increases while the cosmic proton flux decreases with energy. Consequently, cosmic antideuteron production peaks at a laboratory momentum ($p_\text{lab}$) of about 300~\GeVc in \pp interactions.
Currently, the uncertainties in the production of astrophysical antideuteron backgrounds are on the order of a factor of ten, i.e., they range from approximately one-tenth to ten times the predicted values.

In 2025, \NASixtyOne will collect high-statistics \pp data at 300~\GeVc to build a state-of-the-art model with significantly reduced uncertainties to conclusively determine if a conventional astrophysical background can explain the events.
This data set enables the construction of a new state-of-the-art antideuteron model from the source size of \pp interactions and antiproton differential production cross-sections. This model will also be validated by directly measuring the deuteron and antideuteron production cross sections. The significantly improved measurements with \NASixtyOne are critical for evaluating the hypothesized dark matter origin of the AMS-02 antideuteron candidates.

In the post-LS3 period, more \pp data at lower and higher energies can be taken to build a complete model.
The focus will be on understanding the production of antideuterons in the SPS energy range with a 10\% uncertainty level, accompanied by the corresponding precision deuteron and particle correlation results that enable the development of a next-generation light (anti)deuteron formation model.
The same data will also enable the first-time measurement of antihelium-3 production at SPS energies, complementing the high-energy measurements of ALICE.

\subsubsection{Nuclear fragmentation}

Cosmic-ray (CR) propagation characteristics are currently dominated by significant uncertainties in the nuclear fragmentation cross sections, which are approximately 20--30\%. While the precision on recent flux measurements by detectors like AMS-02, CALET, and DAMPE is $<5\%$, it is crucial to perform laboratory measurements of the fragmentation cross sections with the same precision to determine CR transport parameters in the Galaxy. The boron-to-carbon (B/C) flux ratio is the simplest and most well-studied secondary-to-primary ratio. Hence, a precise nuclear fragmentation cross section leading to boron production will aid in studying CR propagation in the Galaxy. 

A first pilot run of carbon fragmentation measurement at 13.5\AGeVc was conducted at \NASixtyOne in 2018, demonstrating that the measurements are possible~\cite{NA61SHINE:2024rzv}. For this type of measurement, the primary $^{208}$Pb is extracted from the SPS and fragmented in collisions with a 160\,mm-long beryllium plate in the H2 beam line. The resulting nuclear fragments of a chosen rigidity are guided to the \NASixtyOne experiment, where the projectile isotopes are identified via a measurement of the particle charge and time-of-flight over a length of approximately $240$\,m. Moreover, data on the fragmentation of nuclei from Li to Si at 12.5\AGeVc were collected at the end of 2024 and are currently being analyzed.
%The collected data were inspired by the interactions listed in Tab.~\ref{tab:ninter}, and they will provide a comprehensive set of cross-sections in the lower triangular region of Fig.~\ref{fig:nucdata_precision}.
High-precision measurements of inelastic and production cross sections necessary for Galactic cosmic-ray data interpretation are also possible, as illustrated in Ref.~\cite{NA61SHINE:2019aip}.
In the future, these measurements of nuclear fragmentation (and inelastic cross sections) with \NASixtyOne can be extended up to Fe and performed at different energies.

\subsection{Neutrino}
%{\itshape \color{blue} Coordinator: Yoshikazu, Eric}

%\textcolor{red}{aim 1 page.   Focus on post-LS3 data collection but still should remind what we collected so far. analysis will be done during LS3. }

\NASixtyOne has executed a broad program of measurements of hadron production for understanding the flux of neutrino beams. In Run~3, high-statistics data sets have been collected on replica targets for T2K and LBNF/DUNE, as well as for hadron interactions on thin solid targets. Concurrently, the flagship physics analyses on Run~2's data were completed and published: 120~\GeVc proton interactions on carbon, with the new forward TPC system providing full forward phase space coverage~\cite{NA61SHINE:2022uxp, NA61SHINE:2023bqo}. 

In 2026, \NASixtyOne plans to begin atmospheric neutrino flux studies with a run of 20~\GeVc protons on a liquid nitrogen target to simulate the dominant interactions of cosmic rays in the atmosphere. These data sets are under current analysis, and we expect the analyses to be completed during LS3.

In Run 4, \NASixtyOne plans to conduct studies of hadron production on nuclear targets using a new very-low-energy (LE) tertiary branch of the H2 beam. These measurements will include both proton- and meson-incident interactions on nitrogen at a momentum below the current H2 beam momentum, down to 2~\GeVc, as well as solid materials, to improve predictions for the flux of neutrinos from atmospheric sources, accelerator beams, and spallation sources. 

The \NASixtyOne-LE beam project has been supported by the leadership of eleven neutrino and muon physics collaborations in a November 2024 letter to CERN's management. From the  CERN side, a technical study was prepared in the framework of PBC-CBWG~\cite{NA61-lowenergy}. Development is currently on hold, awaiting official approval from the relevant CERN committees (SPSC, CERN RB). The new beam will be installed parallel to the existing one, which will remain available for \NASixtyOne and downstream users.

\subsubsection{Physics from the low-energy beam}

\begin{figure}
    \centering
    \includegraphics[width=0.8\linewidth]{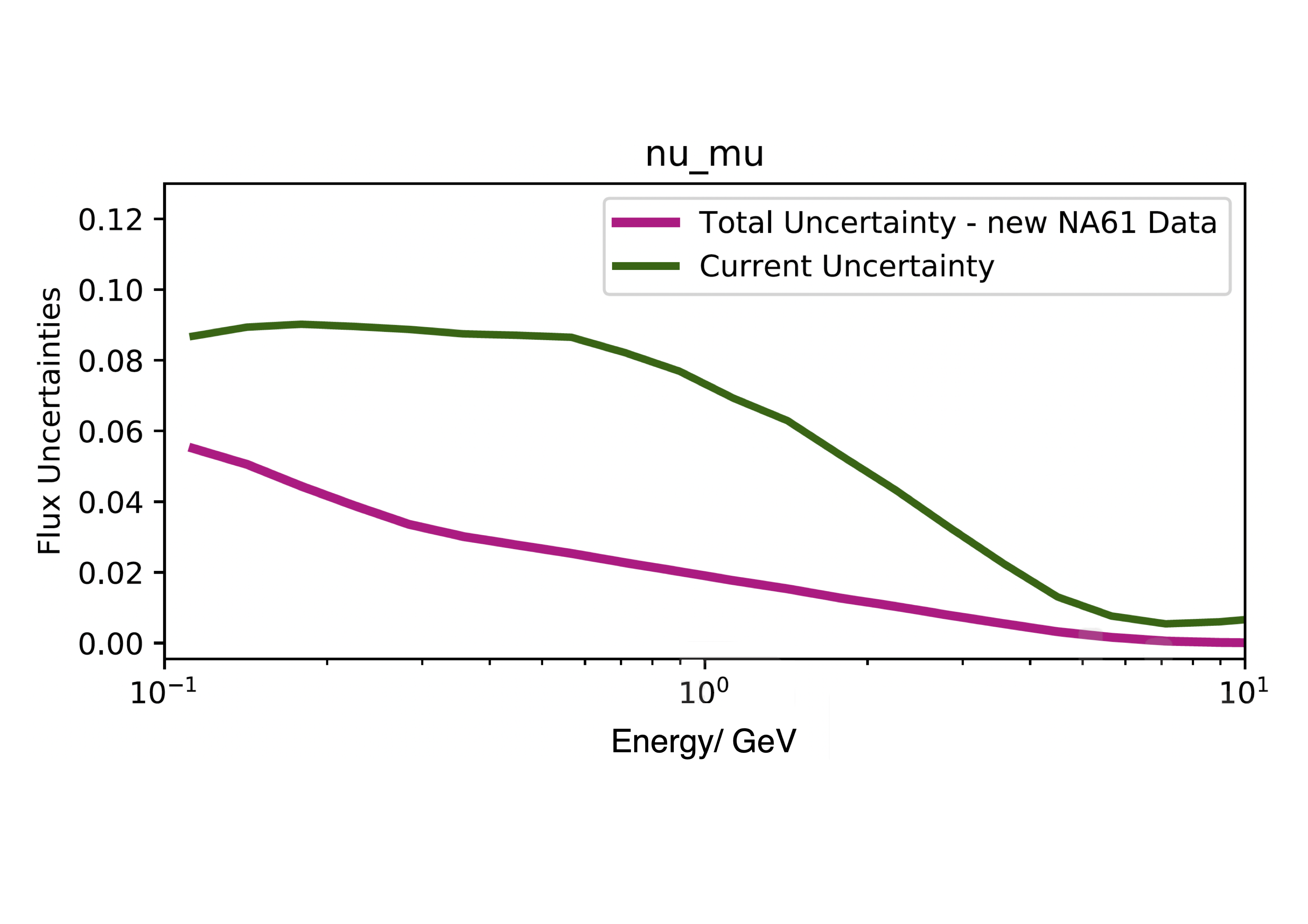}\vspace{-13mm}
    \caption{Atmospheric neutrino flux uncertainties coming from the $p+\mathrm{N}\to \pi^\pm + \mathrm{X}$ process for muon neutrinos. Only the uncertainties due to $< 30$~\GeVc hadronic interactions are considered. The plot shows uncertainty with (pink) and without (green) new low-energy (low-E) NA61/SHINE measurements. Plot by L.~Cook and the Bartol flux group.}
    \label{fig:AtmosphericNuFluxReduction}
\end{figure}

The \NASixtyOne-LE program is summarized in documents~\cite{NA61SHINE:SPSC-P-330-ADD-12,NA61SHINE_SPSC_M_793}. 
Its focus includes proton-nitrogen interactions at energies below 20~\GeVc to improve atmospheric neutrino flux models, low-energy meson interactions on aluminum, water, and iron to improve flux models of long-baseline neutrino beams, and proton-beryllium interactions at 8~\GeVc to reduce the flux uncertainty of Fermilab's Booster Neutrino Beam (BNB).
The measurements for atmospheric neutrinos can potentially reduce the flux uncertainty more than a factor of two for sub-GeV and multi-GeV neutrino fluxes as demonstrated in Fig.~\ref{fig:AtmosphericNuFluxReduction}, which is an important phase-space to study $CP$ symmetry and mass hierarchy with atmospheric neutrinos. 
The measurements for long-baseline neutrino beams, including those of the Hyper-Kamiokande and DUNE experiments, have the potential to reduce the uncertainty in the wrong-sign neutrino flux that originates from unconstrained interactions.
For the BNB flux, there exists a substantial potential to reduce the uncertainty in muon neutrino flux resulting from unconstrained charged pion production in proton-beryllium interactions.

%\begin{itemize}
%\item{A comprehensive study of proton-nitrogen interactions at energies below 20~GeV. These will be used to improve models of atmospheric neutrinos, potentially resulting in improvements to the neutrino flux errors of more than a factor of two.}
%
%\item{Measurements of meson production and rescattering from low-energy interactions on aluminium, water, and iron. These will reduce the current large uncertainties on wrong-sign neutrinos (especially $\nu_e$ and $\bar\nu_e$) that result from interactions outside the target at long-baseline beams and are especially significant backgrounds in precise studies of $CP$ symmetry. These measurements will be geared to the specific needs of the Hyper-K and DUNE experiments.}
%\item {A study of 8~GeV $p+{\rm Be}$ interactions geared toward reducing the flux uncertainty of Fermilab's Booster Neutrino Beam from 7-40\% down to 5-6\% across the entire energy spectrum. This will translate directly into improved $\nu+{\rm Ar}$ cross-section measurements.}
%\end{itemize}

\subsubsection{Physics from the existing high-energy beam}

While most of \NASixtyOne's neutrino-related physics measurements planned for Run 4 will exploit the low-energy beam, it is crucial to preserve the ability to use the current high-energy hadron beam as well. The design of the low-energy beam explicitly allows this, with switchover between the two configurations requiring at most a small number of days.

\NASixtyOne collects high-statistics data sets on replica targets for T2K and LBNF/DUNE during Run~3. However, it is envisioned that the LBNF/DUNE and Hyper-Kamiokande neutrino beams will evolve with new target designs and materials during their expected long operational lives. The current replica-target data sets may not adequately constrain the neutrino flux from these future target configurations. \NASixtyOne will therefore maintain the capability to take new replica-target data sets when new target designs are developed for the neutrino beams. It is not possible to make detailed plans for specific measurements at this time, but they will be proposed as needed in the future by Hyper-Kamiokande and DUNE collaborators who are also members of \NASixtyOne. 
\section{Methodology}
The current configuration of the \NASixtyOne detector is sufficient to continue measurements of hadron production in medium-sized nuclei and to investigate strangeness and multi-strangeness production. However, the proposed measurements for neutrinos and multi-charm correlations require significant detector and beam upgrades respectively.

The existing event rate of the \NASixtyOne detector exceeds 1 kHz, whereas the new measurements demand an increased data-taking rate of up to 10 kHz. To achieve this, the following upgrades are necessary:
\begin{itemize}
\item Construction of a new silicon tracking detector;
\item Development of a new low-energy beamline for neutrino and cosmic-ray physics.
\end{itemize}
\subsection{Construction of a new silicon tracking detector}
%{\itshape \color{blue} Coordinator: Przemysław, Seweryn}
To enable studies of $c-\bar{c}$ correlations, a data-taking rate of 10~kHz or more is necessary~\cite{Gazdzicki:2023niq}. While these rates are easily allowed by the current detector technologies, they exceed the capabilities of the existing \NASixtyOne setup that incorporates large-volume TPCs used for tracking. 
  
To overcome this limitation, we plan to construct a new magnetic spectrometer equipped with a fast silicon tracking detector for primary and decay vertex reconstruction, supplemented with large-area micropattern gaseous detectors (MPGDs) for further tracking to enhance momentum resolution. 
%Figure~\ref{fig:Upgrade:LAST_Outline} schematically depicts the general outline of the new spectrometer. 
According to preliminary Monte Carlo studies, this setup would enable the reconstruction of particle momenta with a resolution of 0.7\% and an acceptance of nearly 40\%. 

The silicon tracking detector will be similar to the existing vertex detector based on ALPIDE sensors, while the technology of the large-area MPGDs is yet to be specified. Based on a Monte Carlo simulation, we expect 10 tracks per cm$^2$ per central Pb+Pb collision and up to $5 \times 10^4$ MIP/cm$^2$/s total charged particle flux in the inner region of the detector and substantially lower in the outer regions.
  
Concerning the timeline, two preliminary scenarios are considered: construction and installation during LS3 to be ready for Run~4, or installation two years into Run~4.
  
%\begin{figure}[ht]
%\begin{center}
%\includegraphics[width=0.5\textwidth]{figures/LAST_Outline.png} 
%\end{center}
%\caption{
%Outline of the proposed new spectrometer enabling studies of $c-\bar{c}$ correlations.
%}
%\label{fig:Upgrade:LAST_Outline}
%\end{figure}
\subsection{Low-energy beamline}
%{\itshape \color{blue} Coordinator: Yoshikazu, Eric}
%\textcolor{blue}{yoshikazu}

\begin{figure}
    \centering
    \includegraphics[width=0.8\linewidth]{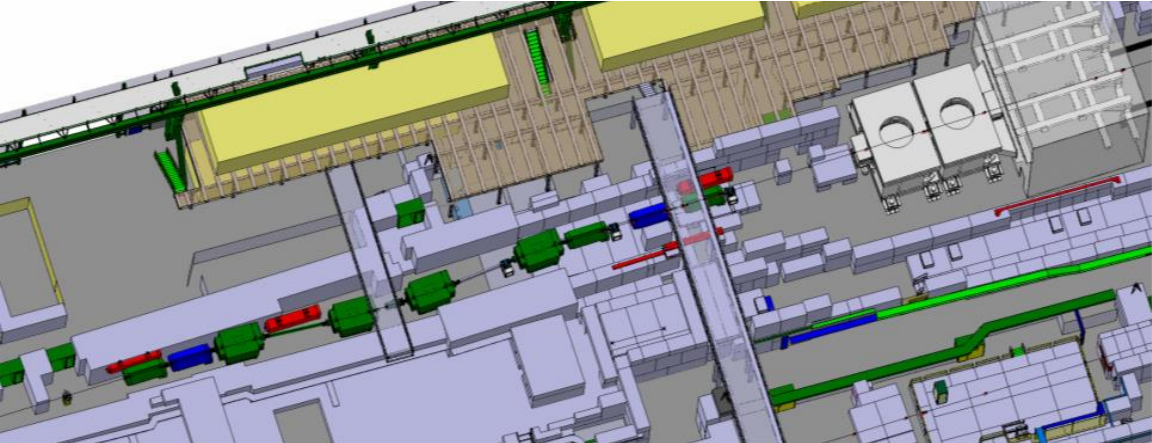}
    \caption{Proposed low-energy branch for  \NASixtyOne. The new magnets for the low-energy particles appear in green, while a rail system would allow the changeover between this low-energy and the normal high-energy configuration. More details on the proposed implementation can be found in Ref.~\cite{NA61-lowenergy}.}
   \label{fig:NA61le}
\end{figure}

The current H2 beamline can provide hadron beams in the range of 13--400~\GeVc.
The goal of the low-E beamline project is to provide hadron beams below this range down to 2~\GeVc. 
The low-E beamline configuration must not prevent standard high-energy operation even after the construction of the new low-E branch of the beamline.
To overcome this challenge, we propose a new high-energy configuration of the H2 line whose beam quality has been carefully assessed and confirmed, as standard operation up to 400~\GeVc remains possible~\cite{NA61-lowenergy}.
The low-E beamline configurations can coexist with the "standard" high-energy configuration as shown in Fig.~\ref{fig:NA61le}, and the H2 line will be able to deliver hadron beams in the range of 2--400~\GeVc.
To keep the pile-up rate at a similar level as was in past NA61/SHINE operations, the beam particle rate needs to be lower than $1.5 \times 10^4$\,Hz~\cite{NA61SHINE:SPSC-P-330-ADD-12}. 
In addition, to collect enough data statistics for spectra measurements within a reasonable beam-time, like two to three weeks per data set, the number of particles has to exceed at least 50\,Hz ($\sim$250 particles per spill).

\section{Readiness and expected challenges}
%{\itshape \color{blue} Coordinator: XYZ}
The \NASixtyOne Collaboration consists of approximately 130 physicists from 30 institutions across 11 countries. The Collaboration has submitted a request for data taking in 2029, marking the start of new measurements. The Collaboration will secure the necessary resources for detector upgrades.

The operation of the experiment during Run 4 will require:
\begin{itemize}
    \item Proton, light ion, and lead beams delivered from the CERN SPS for several months per year, as \NASixtyOne operates in parallel with other fixed-target experiments and the LHC;
\item CERN IT resources for data processing and storage;
\item Detector operation and maintenance, including data acquisition and expert shifts, require approximately 500~kCHF per year, which the Collaboration will cover;
\item Construction of the low-energy beamline to support new measurements.
\end{itemize}
To fully exploit the physics opportunities offered by nuclear beams, delivering a wider variety of ion species for different experiments would be highly beneficial. The proposed addition of a second heavy-ion source and upgrades to the heavy-ion complex would significantly enhance the scientific potential, not only for \NASixtyOne but for all heavy-ion experiments at CERN.

\section{Recommendations for support by the ESPP}
The \NASixtyOne Collaboration requests a recommendation from the European Strategy for Particle Physics to support data-taking measurements for:
\begin{itemize}
    \item Continuing the exploration of the transition to the quark-gluon plasma with increasing system size using light-ion collisions;
    \item Conducting measurements for neutrino and cosmic-ray physics, particularly focusing on developing the low-energy beamline and comprehensive, large-acceptance hadron measurements.
\end{itemize}
In addition, support should continue for fixed-target programs at CERN SPS, RHIC, and FAIR, which address unique and complementary aspects of QCD.
%\bibliographystyle{unsrt}
%\bibliography{main}
%\newpage
%\input{info}
\newpage
{
\section*{Acknowledgments}
We would like to thank the CERN EP, BE, HSE and EN Departments for the
strong support of NA61/SHINE.

This work was supported by
the Hungarian Scientific Research Fund (grant NKFIH 138136\slash137812\slash138152 and TKP2021-NKTA-64),
the Polish Ministry of Science and Higher Education
(DIR\slash WK\slash\-2016\slash 2017\slash\-10-1, WUT ID-UB), the National Science Centre Poland (grants
2014\slash 14\slash E\slash ST2\slash 00018, %AR, settled
2016\slash 21\slash D\slash ST2\slash 01983, %MMP, settled
2017\slash 25\slash N\slash ST2\slash 02575, %AT, settled
2018\slash 29\slash N\slash ST2\slash 02595, %AM, completed, not settled
2018\slash 30\slash A\slash ST2\slash 00226, %MG, in progress
2018\slash 31\slash G\slash ST2\slash 03910, %SK, in progress
2020\slash 39\slash O\slash ST2\slash 00277, %MR, in progress
2021\slash 43\slash P\slash ST2\slash 03319), %LT, in progress
the Norwegian Financial Mechanism 2014--2021 (grant 2019\slash 34\slash H\slash ST2\slash 00585),
the Polish Minister of Education and Science (contract No. 2021\slash WK\slash 10),
%the Russian Science Foundation (grant 17-72-20045),
%the Russian Academy of Science and the
%Russian Foundation for Basic Research (grants 08-02-00018, 09-02-00664 and 12-02-91503-CERN),
%the Russian Foundation for Basic Research (RFBR) funding within the research project no. 18-02-40086,
%the Ministry of Science and Higher Education of the Russian Federation, Project "Fundamental properties of elementary particles and cosmology" No 0723-2020-0041,
the European Union's Horizon 2020 research and innovation programme under grant agreement No. 871072,
the Ministry of Education, Culture, Sports,
Science and Tech\-no\-lo\-gy, Japan, Grant-in-Aid for Sci\-en\-ti\-fic
Research (grants 18071005, 19034011, 19740162, 20740160 and 20039012, 22H04\\943), 
the German Research Foundation DFG (grants GA\,1480\slash8-1 and project 426579465),
the Bulgarian Ministry of Education and Science within the National
Roadmap for Research Infrastructures 2020--2027, contract No. D01-374/18.12.2020,
Serbian Ministry of Science, Technological Development and Innovation (grant
OI171002), Swiss Nationalfonds Foundation (grant 200020\-117913/1),
ETH Research Grant TH-01\,07-3, National Science Foundation grant
PHY-2013228 and the Fermi National Accelerator Laboratory (Fermilab),
a U.S. Department of Energy, Office of Science, HEP User Facility
managed by Fermi Research Alliance, LLC (FRA), acting under Contract
No. DE-AC02-07CH11359 and the IN2P3-CNRS (France).\\

% The data used in this paper were collected before February 2022.

}
\newpage
\begin{multicols}{2}
\bibliographystyle{include/na61Utphys}
%{\footnotesize\raggedright
{\footnotesize
%\bibliography{include/na61References}
\bibliography{references}
}
\end{multicols}
\newpage
{\Large The \NASixtyOne Collaboration}
\bigskip
\begin{sloppypar}
% based on XML DB with time Mon Jul  7 16:27:28 2025
% Authors in alphabetical order.

\noindent
{H.\;Adhikary~\href{https://orcid.org/0000-0002-5746-1268}{\includegraphics[height=1.7ex]{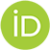}}\textsuperscript{\,11}},
{P.\;Adrich~\href{https://orcid.org/0000-0002-7019-5451}{\includegraphics[height=1.7ex]{orcid-logo.png}}\textsuperscript{\,13}},
{K.K.\;Allison~\href{https://orcid.org/0000-0002-3494-9383}{\includegraphics[height=1.7ex]{orcid-logo.png}}\textsuperscript{\,24}},
{N.\;Amin~\href{https://orcid.org/0009-0004-7572-3817}{\includegraphics[height=1.7ex]{orcid-logo.png}}\textsuperscript{\,4}},
{E.V.\;Andronov~\href{https://orcid.org/0000-0003-0437-9292}{\includegraphics[height=1.7ex]{orcid-logo.png}}\textsuperscript{\,21}},
{I.-C.\;Arsene~\href{https://orcid.org/0000-0003-2316-9565}{\includegraphics[height=1.7ex]{orcid-logo.png}}\textsuperscript{\,10}},
{M.\;Bajda~\href{https://orcid.org/0009-0005-8859-1099}{\includegraphics[height=1.7ex]{orcid-logo.png}}\textsuperscript{\,14}},
{Y.\;Balkova~\href{https://orcid.org/0000-0002-6957-573X}{\includegraphics[height=1.7ex]{orcid-logo.png}}\textsuperscript{\,13}},
{D.\;Battaglia~\href{https://orcid.org/0000-0002-5283-0992}{\includegraphics[height=1.7ex]{orcid-logo.png}}\textsuperscript{\,23}},
{A.\;Bazgir~\href{https://orcid.org/0000-0003-0358-0576}{\includegraphics[height=1.7ex]{orcid-logo.png}}\textsuperscript{\,11}},
{S.\;Bhosale~\href{https://orcid.org/0000-0001-5709-4747}{\includegraphics[height=1.7ex]{orcid-logo.png}}\textsuperscript{\,12}},
{M.\;Bielewicz~\href{https://orcid.org/0000-0001-8267-4874}{\includegraphics[height=1.7ex]{orcid-logo.png}}\textsuperscript{\,13}},
{A.\;Blondel~\href{https://orcid.org/0000-0002-1597-8859}{\includegraphics[height=1.7ex]{orcid-logo.png}}\textsuperscript{\,3}},
{M.\;Bogomilov~\href{https://orcid.org/0000-0001-7738-2041}{\includegraphics[height=1.7ex]{orcid-logo.png}}\textsuperscript{\,2}},
{Y.\;Bondar~\href{https://orcid.org/0000-0003-2773-9668}{\includegraphics[height=1.7ex]{orcid-logo.png}}\textsuperscript{\,11}},
{W.\;Bryli\'nski~\href{https://orcid.org/0000-0002-3457-6601}{\includegraphics[height=1.7ex]{orcid-logo.png}}\textsuperscript{\,19}},
{J.\;Brzychczyk~\href{https://orcid.org/0000-0001-5320-6748}{\includegraphics[height=1.7ex]{orcid-logo.png}}\textsuperscript{\,14}},
{M.\;Buryakov~\href{https://orcid.org/0009-0008-2394-4967}{\includegraphics[height=1.7ex]{orcid-logo.png}}\textsuperscript{\,20}},
{A.F.\;Camino\textsuperscript{\,26}},
{Y.D.\;Chandak~\href{https://orcid.org/0009-0009-2080-566X}{\includegraphics[height=1.7ex]{orcid-logo.png}}\textsuperscript{\,24}},
{M.\;Csan\'ad~\href{https://orcid.org/0000-0002-3154-6925}{\includegraphics[height=1.7ex]{orcid-logo.png}}\textsuperscript{\,6}},
{J.\;Cybowska~\href{https://orcid.org/0000-0003-2568-3664}{\includegraphics[height=1.7ex]{orcid-logo.png}}\textsuperscript{\,19}},
{T.\;Czopowicz~\href{https://orcid.org/0000-0003-1908-2977}{\includegraphics[height=1.7ex]{orcid-logo.png}}\textsuperscript{\,13}},
{C.\;Dalmazzone~\href{https://orcid.org/0000-0001-6945-5845}{\includegraphics[height=1.7ex]{orcid-logo.png}}\textsuperscript{\,3}},
{N.\;Davis~\href{https://orcid.org/0000-0003-3047-6854}{\includegraphics[height=1.7ex]{orcid-logo.png}}\textsuperscript{\,12}},
{A.\;Dmitriev~\href{https://orcid.org/0000-0001-7853-0173}{\includegraphics[height=1.7ex]{orcid-logo.png}}\textsuperscript{\,20}},
{P.~von\;Doetinchem~\href{https://orcid.org/0000-0002-7801-3376}{\includegraphics[height=1.7ex]{orcid-logo.png}}\textsuperscript{\,25}},
{W.\;Dominik~\href{https://orcid.org/0000-0001-7444-9239}{\includegraphics[height=1.7ex]{orcid-logo.png}}\textsuperscript{\,17}},
{J.\;Dumarchez~\href{https://orcid.org/0000-0002-9243-4425}{\includegraphics[height=1.7ex]{orcid-logo.png}}\textsuperscript{\,3}},
{R.\;Engel~\href{https://orcid.org/0000-0003-2924-8889}{\includegraphics[height=1.7ex]{orcid-logo.png}}\textsuperscript{\,4}},
{G.A.\;Feofilov~\href{https://orcid.org/0000-0003-3700-8623}{\includegraphics[height=1.7ex]{orcid-logo.png}}\textsuperscript{\,21}},
{L.\;Fields~\href{https://orcid.org/0000-0001-8281-3686}{\includegraphics[height=1.7ex]{orcid-logo.png}}\textsuperscript{\,23}},
{Z.\;Fodor~\href{https://orcid.org/0000-0003-2519-5687}{\includegraphics[height=1.7ex]{orcid-logo.png}}\textsuperscript{\,5,18}},
{M.\;Friend~\href{https://orcid.org/0000-0003-4660-4670}{\includegraphics[height=1.7ex]{orcid-logo.png}}\textsuperscript{\,7}},
{M.\;Ga\'zdzicki~\href{https://orcid.org/0000-0002-6114-8223}{\includegraphics[height=1.7ex]{orcid-logo.png}}\textsuperscript{\,11}},
{K.E.\;Gollwitzer\textsuperscript{\,22}},
{O.\;Golosov~\href{https://orcid.org/0000-0001-6562-2925}{\includegraphics[height=1.7ex]{orcid-logo.png}}\textsuperscript{\,21}},
{V.\;Golovatyuk~\href{https://orcid.org/0009-0006-5201-0990}{\includegraphics[height=1.7ex]{orcid-logo.png}}\textsuperscript{\,20}},
{M.\;Golubeva~\href{https://orcid.org/0009-0003-4756-2449}{\includegraphics[height=1.7ex]{orcid-logo.png}}\textsuperscript{\,21}},
{K.\;Grebieszkow~\href{https://orcid.org/0000-0002-6754-9554}{\includegraphics[height=1.7ex]{orcid-logo.png}}\textsuperscript{\,19}},
{F.\;Guber~\href{https://orcid.org/0000-0001-8790-3218}{\includegraphics[height=1.7ex]{orcid-logo.png}}\textsuperscript{\,21}},
{P.G.\;Hurh~\href{https://orcid.org/0000-0002-9024-5399}{\includegraphics[height=1.7ex]{orcid-logo.png}}\textsuperscript{\,22}},
{S.\;Ilieva~\href{https://orcid.org/0000-0001-9204-2563}{\includegraphics[height=1.7ex]{orcid-logo.png}}\textsuperscript{\,2}},
{A.\;Ivashkin~\href{https://orcid.org/0000-0003-4595-5866}{\includegraphics[height=1.7ex]{orcid-logo.png}}\textsuperscript{\,21}},
{N.\;Karpushkin~\href{https://orcid.org/0000-0001-5513-9331}{\includegraphics[height=1.7ex]{orcid-logo.png}}\textsuperscript{\,21}},
{M.\;Kie{\l}bowicz~\href{https://orcid.org/0000-0002-4403-9201}{\includegraphics[height=1.7ex]{orcid-logo.png}}\textsuperscript{\,12}},
{V.A.\;Kireyeu~\href{https://orcid.org/0000-0002-5630-9264}{\includegraphics[height=1.7ex]{orcid-logo.png}}\textsuperscript{\,20}},
{R.\;Kolesnikov~\href{https://orcid.org/0009-0006-4224-1058}{\includegraphics[height=1.7ex]{orcid-logo.png}}\textsuperscript{\,20}},
{D.\;Kolev~\href{https://orcid.org/0000-0002-9203-4739}{\includegraphics[height=1.7ex]{orcid-logo.png}}\textsuperscript{\,2}},
{Y.\;Koshio~\href{https://orcid.org/0000-0003-0437-8505}{\includegraphics[height=1.7ex]{orcid-logo.png}}\textsuperscript{\,8}},
{S.\;Kowalski~\href{https://orcid.org/0000-0001-9888-4008}{\includegraphics[height=1.7ex]{orcid-logo.png}}\textsuperscript{\,16}},
{B.\;Koz{\l}owski~\href{https://orcid.org/0000-0001-8442-2320}{\includegraphics[height=1.7ex]{orcid-logo.png}}\textsuperscript{\,19}},
{A.\;Krasnoperov~\href{https://orcid.org/0000-0002-1425-2861}{\includegraphics[height=1.7ex]{orcid-logo.png}}\textsuperscript{\,20}},
{W.\;Kucewicz~\href{https://orcid.org/0000-0002-2073-711X}{\includegraphics[height=1.7ex]{orcid-logo.png}}\textsuperscript{\,15}},
{M.\;Kuchowicz~\href{https://orcid.org/0000-0003-3174-585X}{\includegraphics[height=1.7ex]{orcid-logo.png}}\textsuperscript{\,18}},
{P.\;Lasko~\href{https://orcid.org/0000-0003-1110-9522}{\includegraphics[height=1.7ex]{orcid-logo.png}}\textsuperscript{\,14}},
{A.\;L\'aszl\'o~\href{https://orcid.org/0000-0003-2712-6968}{\includegraphics[height=1.7ex]{orcid-logo.png}}\textsuperscript{\,5}},
{M.\;Lewicki~\href{https://orcid.org/0000-0002-8972-3066}{\includegraphics[height=1.7ex]{orcid-logo.png}}\textsuperscript{\,12}},
{G.\;Lykasov~\href{https://orcid.org/0000-0002-1544-6959}{\includegraphics[height=1.7ex]{orcid-logo.png}}\textsuperscript{\,20}},
{J.R.\;Lyon~\href{https://orcid.org/0009-0003-2579-8821}{\includegraphics[height=1.7ex]{orcid-logo.png}}\textsuperscript{\,25}},
{V.V.\;Lyubushkin~\href{https://orcid.org/0000-0003-0136-233X}{\includegraphics[height=1.7ex]{orcid-logo.png}}\textsuperscript{\,20}},
{M.\;Ma\'ckowiak-Paw{\l}owska~\href{https://orcid.org/0000-0003-3954-6329}{\includegraphics[height=1.7ex]{orcid-logo.png}}\textsuperscript{\,19}},
{B.\;Maksiak~\href{https://orcid.org/0000-0002-7950-2307}{\includegraphics[height=1.7ex]{orcid-logo.png}}\textsuperscript{\,13}},
{A.I.\;Malakhov~\href{https://orcid.org/0000-0001-8569-8409}{\includegraphics[height=1.7ex]{orcid-logo.png}}\textsuperscript{\,20}},
{A.\;Marcinek~\href{https://orcid.org/0000-0001-9922-743X}{\includegraphics[height=1.7ex]{orcid-logo.png}}\textsuperscript{\,12}},
{A.D.\;Marino~\href{https://orcid.org/0000-0002-1709-538X}{\includegraphics[height=1.7ex]{orcid-logo.png}}\textsuperscript{\,24}},
{T.\;Matulewicz~\href{https://orcid.org/0000-0003-2098-1216}{\includegraphics[height=1.7ex]{orcid-logo.png}}\textsuperscript{\,17}},
{V.\;Matveev~\href{https://orcid.org/0000-0002-2745-5908}{\includegraphics[height=1.7ex]{orcid-logo.png}}\textsuperscript{\,20}},
{G.L.\;Melkumov~\href{https://orcid.org/0009-0004-2074-6755}{\includegraphics[height=1.7ex]{orcid-logo.png}}\textsuperscript{\,20}},
{A.\;Merzlaya~\href{https://orcid.org/0000-0002-6553-2783}{\includegraphics[height=1.7ex]{orcid-logo.png}}\textsuperscript{\,10}},
{{\L}.\;Mik~\href{https://orcid.org/0000-0003-2712-6861}{\includegraphics[height=1.7ex]{orcid-logo.png}}\textsuperscript{\,15}},
{S.\;Morozov~\href{https://orcid.org/0000-0002-6748-7277}{\includegraphics[height=1.7ex]{orcid-logo.png}}\textsuperscript{\,21}},
{Y.\;Nagai~\href{https://orcid.org/0000-0002-1792-5005}{\includegraphics[height=1.7ex]{orcid-logo.png}}\textsuperscript{\,6}},
{R.\;Nagy~\href{https://orcid.org/0009-0004-4274-1832}{\includegraphics[height=1.7ex]{orcid-logo.png}}\textsuperscript{\,5}},
{T.\;Nakadaira~\href{https://orcid.org/0000-0003-4327-7598}{\includegraphics[height=1.7ex]{orcid-logo.png}}\textsuperscript{\,7}},
{S.\;Nishimori~\href{https://orcid.org/~0000-0002-1820-0938}{\includegraphics[height=1.7ex]{orcid-logo.png}}\textsuperscript{\,7}},
{A.\;Olivier~\href{https://orcid.org/0000-0003-4261-8303}{\includegraphics[height=1.7ex]{orcid-logo.png}}\textsuperscript{\,23}},
{V.\;Ozvenchuk~\href{https://orcid.org/0000-0002-7821-7109}{\includegraphics[height=1.7ex]{orcid-logo.png}}\textsuperscript{\,12}},
{O.\;Panova~\href{https://orcid.org/0000-0001-5039-7788}{\includegraphics[height=1.7ex]{orcid-logo.png}}\textsuperscript{\,11}},
{V.\;Paolone~\href{https://orcid.org/0000-0003-2162-0957}{\includegraphics[height=1.7ex]{orcid-logo.png}}\textsuperscript{\,26}},
{I.\;Pidhurskyi~\href{https://orcid.org/0000-0001-9916-9436}{\includegraphics[height=1.7ex]{orcid-logo.png}}\textsuperscript{\,11}},
{R.\;P{\l}aneta~\href{https://orcid.org/0000-0001-8007-8577}{\includegraphics[height=1.7ex]{orcid-logo.png}}\textsuperscript{\,14}},
{P.\;Podlaski~\href{https://orcid.org/0000-0002-0232-9841}{\includegraphics[height=1.7ex]{orcid-logo.png}}\textsuperscript{\,17}},
{B.A.\;Popov~\href{https://orcid.org/0000-0001-5416-9301}{\includegraphics[height=1.7ex]{orcid-logo.png}}\textsuperscript{\,20,3}},
{B.\;P\'orfy~\href{https://orcid.org/0000-0001-5724-9737}{\includegraphics[height=1.7ex]{orcid-logo.png}}\textsuperscript{\,5,6}},
{D.S.\;Prokhorova~\href{https://orcid.org/0000-0003-3726-9196}{\includegraphics[height=1.7ex]{orcid-logo.png}}\textsuperscript{\,21}},
{D.\;Pszczel~\href{https://orcid.org/0000-0002-4697-6688}{\includegraphics[height=1.7ex]{orcid-logo.png}}\textsuperscript{\,13}},
{S.\;Pu{\l}awski~\href{https://orcid.org/0000-0003-1982-2787}{\includegraphics[height=1.7ex]{orcid-logo.png}}\textsuperscript{\,16}},
{L.\;Ren~\href{https://orcid.org/0000-0003-1709-7673}{\includegraphics[height=1.7ex]{orcid-logo.png}}\textsuperscript{\,24}},
{V.Z.\;Reyna~Ortiz~\href{https://orcid.org/0000-0002-7026-8198}{\includegraphics[height=1.7ex]{orcid-logo.png}}\textsuperscript{\,11}},
{D.\;R\"ohrich\textsuperscript{\,9}},
{M.\;Roth~\href{https://orcid.org/0000-0003-1281-4477}{\includegraphics[height=1.7ex]{orcid-logo.png}}\textsuperscript{\,4}},
{{\L}.\;Rozp{\l}ochowski~\href{https://orcid.org/0000-0003-3680-6738}{\includegraphics[height=1.7ex]{orcid-logo.png}}\textsuperscript{\,12}},
{M.\;Rumyantsev~\href{https://orcid.org/0000-0001-8233-2030}{\includegraphics[height=1.7ex]{orcid-logo.png}}\textsuperscript{\,20}},
{A.\;Rustamov~\href{https://orcid.org/0000-0001-8678-6400}{\includegraphics[height=1.7ex]{orcid-logo.png}}\textsuperscript{\,1}},
{M.\;Rybczynski~\href{https://orcid.org/0000-0002-3638-3766}{\includegraphics[height=1.7ex]{orcid-logo.png}}\textsuperscript{\,11}},
{A.\;Rybicki~\href{https://orcid.org/0000-0003-3076-0505}{\includegraphics[height=1.7ex]{orcid-logo.png}}\textsuperscript{\,12}},
{D.\;Rybka~\href{https://orcid.org/0000-0002-9924-6398}{\includegraphics[height=1.7ex]{orcid-logo.png}}\textsuperscript{\,13}},
{K.\;Sakashita~\href{https://orcid.org/0000-0003-2602-7837}{\includegraphics[height=1.7ex]{orcid-logo.png}}\textsuperscript{\,7}},
{K.\;Schmidt~\href{https://orcid.org/0000-0002-0903-5790}{\includegraphics[height=1.7ex]{orcid-logo.png}}\textsuperscript{\,16}},
{P.\;Seyboth~\href{https://orcid.org/0000-0002-4821-6105}{\includegraphics[height=1.7ex]{orcid-logo.png}}\textsuperscript{\,11}},
{U.A.\;Shah~\href{https://orcid.org/0000-0002-9315-1304}{\includegraphics[height=1.7ex]{orcid-logo.png}}\textsuperscript{\,11}},
{Y.\;Shiraishi~\href{https://orcid.org/0000-0002-0132-3923}{\includegraphics[height=1.7ex]{orcid-logo.png}}\textsuperscript{\,8}},
{A.\;Shukla~\href{https://orcid.org/0000-0003-3839-7229}{\includegraphics[height=1.7ex]{orcid-logo.png}}\textsuperscript{\,25}},
{M.\;S{\l}odkowski~\href{https://orcid.org/0000-0003-0463-2753}{\includegraphics[height=1.7ex]{orcid-logo.png}}\textsuperscript{\,19}},
{P.\;Staszel~\href{https://orcid.org/0000-0003-4002-1626}{\includegraphics[height=1.7ex]{orcid-logo.png}}\textsuperscript{\,14}},
{G.\;Stefanek~\href{https://orcid.org/0000-0001-6656-9177}{\includegraphics[height=1.7ex]{orcid-logo.png}}\textsuperscript{\,11}},
{J.\;Stepaniak~\href{https://orcid.org/0000-0003-2064-9870}{\includegraphics[height=1.7ex]{orcid-logo.png}}\textsuperscript{\,13}},
{{\L}.\;\'Swiderski~\href{https://orcid.org/0000-0001-5857-2085}{\includegraphics[height=1.7ex]{orcid-logo.png}}\textsuperscript{\,13}},
{J.\;Szewi\'nski~\href{https://orcid.org/0000-0003-2981-9303}{\includegraphics[height=1.7ex]{orcid-logo.png}}\textsuperscript{\,13}},
{R.\;Szukiewicz~\href{https://orcid.org/0000-0002-1291-4040}{\includegraphics[height=1.7ex]{orcid-logo.png}}\textsuperscript{\,18}},
{A.\;Taranenko~\href{https://orcid.org/0000-0003-1737-4474}{\includegraphics[height=1.7ex]{orcid-logo.png}}\textsuperscript{\,21}},
{A.\;Tefelska~\href{https://orcid.org/0000-0002-6069-4273}{\includegraphics[height=1.7ex]{orcid-logo.png}}\textsuperscript{\,19}},
{D.\;Tefelski~\href{https://orcid.org/0000-0003-0802-2290}{\includegraphics[height=1.7ex]{orcid-logo.png}}\textsuperscript{\,19}},
{V.\;Tereshchenko\textsuperscript{\,20}},
{R.\;Tsenov~\href{https://orcid.org/0000-0002-1330-8640}{\includegraphics[height=1.7ex]{orcid-logo.png}}\textsuperscript{\,2}},
{L.\;Turko~\href{https://orcid.org/0000-0002-5474-8650}{\includegraphics[height=1.7ex]{orcid-logo.png}}\textsuperscript{\,18}},
{T.S.\;Tveter~\href{https://orcid.org/0009-0003-7140-8644}{\includegraphics[height=1.7ex]{orcid-logo.png}}\textsuperscript{\,10}},
{M.\;Unger~\href{https://orcid.org/0000-0002-7651-0272}{\includegraphics[height=1.7ex]{orcid-logo.png}}\textsuperscript{\,4}},
{M.\;Urbaniak~\href{https://orcid.org/0000-0002-9768-030X}{\includegraphics[height=1.7ex]{orcid-logo.png}}\textsuperscript{\,16}},
{D.\;Veberi\v{c}~\href{https://orcid.org/0000-0003-2683-1526}{\includegraphics[height=1.7ex]{orcid-logo.png}}\textsuperscript{\,4}},
{O.\;Vitiuk~\href{https://orcid.org/0000-0002-9744-3937}{\includegraphics[height=1.7ex]{orcid-logo.png}}\textsuperscript{\,18}},
{A.\;Wickremasinghe~\href{https://orcid.org/0000-0002-5325-0455}{\includegraphics[height=1.7ex]{orcid-logo.png}}\textsuperscript{\,22}},
{K.\;Witek~\href{https://orcid.org/0009-0004-6699-1895}{\includegraphics[height=1.7ex]{orcid-logo.png}}\textsuperscript{\,15}},
{K.\;W\'ojcik~\href{https://orcid.org/0000-0002-8315-9281}{\includegraphics[height=1.7ex]{orcid-logo.png}}\textsuperscript{\,16}},
{O.\;Wyszy\'nski~\href{https://orcid.org/0000-0002-6652-0450}{\includegraphics[height=1.7ex]{orcid-logo.png}}\textsuperscript{\,11}},
{A.\;Zaitsev~\href{https://orcid.org/0000-0003-4711-9925}{\includegraphics[height=1.7ex]{orcid-logo.png}}\textsuperscript{\,20}},
{E.\;Zherebtsova~\href{https://orcid.org/0000-0002-1364-0969}{\includegraphics[height=1.7ex]{orcid-logo.png}}\textsuperscript{\,18}},
{E.D.\;Zimmerman~\href{https://orcid.org/0000-0002-6394-6659}{\includegraphics[height=1.7ex]{orcid-logo.png}}\textsuperscript{\,24}}, and
{A.\;Zviagina~\href{https://orcid.org/0009-0007-5211-6493}{\includegraphics[height=1.7ex]{orcid-logo.png}}\textsuperscript{\,21}}

\end{sloppypar}
% based on XML DB with time Mon Jul  7 16:27:28 2025
% Institutes in alphabetical order.

\noindent
\textsuperscript{1}~National Nuclear Research Center, Baku, Azerbaijan\\
\textsuperscript{2}~Faculty of Physics, University of Sofia, Sofia, Bulgaria\\
\textsuperscript{3}~LPNHE, Sorbonne University, CNRS/IN2P3, Paris, France\\
\textsuperscript{4}~Karlsruhe Institute of Technology, Karlsruhe, Germany\\
\textsuperscript{5}~HUN-REN Wigner Research Centre for Physics, Budapest, Hungary\\
\textsuperscript{6}~E\"otv\"os Lor\'and University, Budapest, Hungary\\
\textsuperscript{7}~Institute for Particle and Nuclear Studies, Tsukuba, Japan\\
\textsuperscript{8}~Okayama University, Japan\\
\textsuperscript{9}~University of Bergen, Bergen, Norway\\
\textsuperscript{10}~University of Oslo, Oslo, Norway\\
\textsuperscript{11}~Jan Kochanowski University, Kielce, Poland\\
\textsuperscript{12}~Institute of Nuclear Physics, Polish Academy of Sciences, Cracow, Poland\\
\textsuperscript{13}~National Centre for Nuclear Research, Warsaw, Poland\\
\textsuperscript{14}~Jagiellonian University, Cracow, Poland\\
\textsuperscript{15}~AGH - University of Krakow, Poland\\
\textsuperscript{16}~University of Silesia, Katowice, Poland\\
\textsuperscript{17}~University of Warsaw, Warsaw, Poland\\
\textsuperscript{18}~University of Wroc{\l}aw,  Wroc{\l}aw, Poland\\
\textsuperscript{19}~Warsaw University of Technology, Warsaw, Poland\\
\textsuperscript{20}~Joint Institute for Nuclear Research, Dubna, International Organization\\
\textsuperscript{21}~Affiliated with an institution formerly covered by a cooperation agreement with CERN\\
\textsuperscript{22}~Fermilab, Batavia, USA\\
\textsuperscript{23}~University of Notre Dame, Notre Dame, USA\\
\textsuperscript{24}~University of Colorado, Boulder, USA\\
\textsuperscript{25}~University of Hawaii at Manoa, Honolulu, USA\\
\textsuperscript{26}~University of Pittsburgh, Pittsburgh, USA\\

\end{document}